\documentclass[%
aip,
amsmath,amssymb,
reprint,%
]{revtex4-1}

\usepackage{graphicx}
\usepackage{dcolumn}
\usepackage{bm}
\usepackage[utf8]{inputenc}
\usepackage[T1]{fontenc}
\usepackage{mathptmx}

\begin{document}

\onecolumngrid
\noindent This is the pre-peer reviewed version of the following article: Single‐Material Graphene Thermocouples. \textit{Adv. Funct. Mater.} 2020, 2000574, which has been published in final form at https://doi.org/10.1002/adfm.202000574. This article may be used for non-commercial purposes in accordance with Wiley Terms and Conditions for Use of Self-Archived Versions.

\title{Single-Material Graphene Thermocouples} 



\author{A. Harzheim}
\email[]{achim.harzheim@materials.ox.ac.uk}
\affiliation{Department of Materials, University of Oxford, Oxford, United Kingdom}
\author{F. Koenemann}
\affiliation{IBM Research - Zurich, 8803 Rueschlikon, Switzerland}
\author{B. Gotsmann}
\affiliation{IBM Research - Zurich, 8803 Rueschlikon, Switzerland}
\author{H. van der Zant}
\affiliation{Kavli Institute of Nanoscience, Delft University of Technology, Delft, Netherlands}
\author{P. Gehring}
\email[]{p.gehring@tudelft.nl}
\altaffiliation[Current Address:]{IMEC, Kapeldreef 75, 3001 Leuven, Belgium}
\affiliation{Kavli Institute of Nanoscience, Delft University of Technology, Delft, Netherlands}



\begin{abstract}
On-chip temperature sensing on a micro- to nanometer scale is becoming more desirable as the complexity of nanodevices and size requirements increase and with it the challenges in thermal probing and management. This highlights the need for scalable and reliable temperature sensors which have the potential to be incorporated into current and future device structures. Here, we show that U-shaped graphene stripes consisting of one wide and one narrow leg form a single material thermocouple that can function as a self-powering temperature sensor. We find that the graphene thermocouples increase in sensitivity with a decrease in leg width, due to a change in the Seebeck coefficient, which is in agreement with our previous findings and report a maximum sensitivity of $\Delta S \approx$ 39 $\mathrm{\mu}$V/K.
\end{abstract}


\maketitle 

\begin{figure}
	\includegraphics[width=\linewidth]{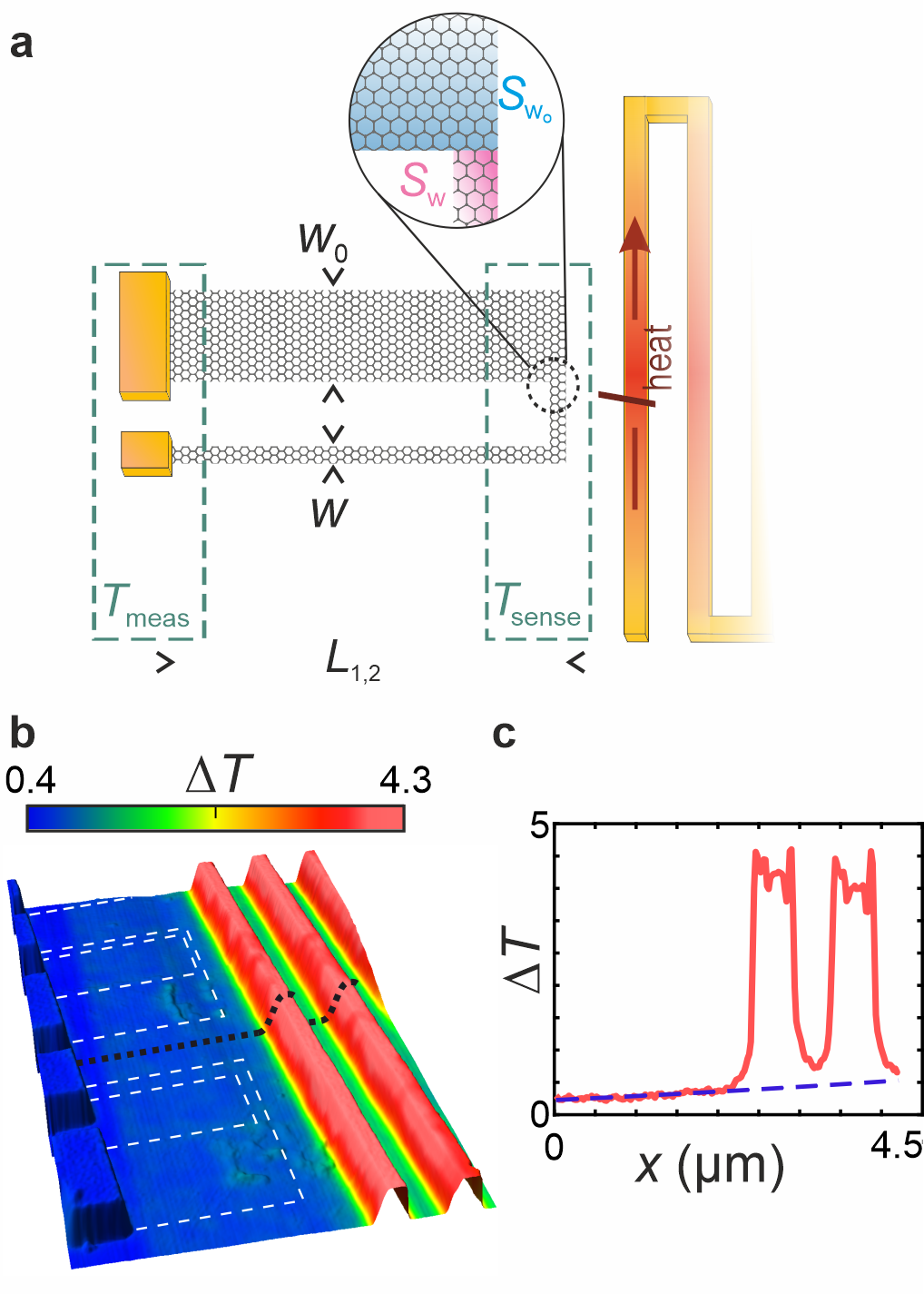}
	\caption{\label{Fig:1:Schematics}a) Schematic depiction of the device structure. b) Temperature distribution of a typical operating device measured using SThM at a heater power of 6.4 mW. The white dashed lines indicate the position of the graphene thermocouples and the black dashed line indicates the line cut shown in (c). c) Temperature profile line extracted from (b) in pink and the fit along the substrate (blue dashed line).}	
\end{figure}

Complex electronic devices and circuits rely on thermal sensors incorporated into the structure to give input to the power and thermal management system \cite{Skadron2003,Gunther2001}. In order to avoid hot spots, built-in temperature sensors are distributed along critical points to monitor the temperature and provide feedback to the control system \cite{Semenov2006,Mukherjee2006}. This allows for the redistribution of the thermal load through spot cooling or load distribution, e.g. among different computing cores, enabling a longer device lifetime and saving energy. Ideally, these temperature sensors need to have a small footprint, high accuracy, consume a minimum amount of power and be compatible with established nanofabrication techniques. Today, resistors or diodes are often used for on-chip temperature sensing, with diodes providing a high sensitivity. However, since p-n junction diodes are made from doped semiconductors they can be subject to fluctuations in the doping and have to be calibrated individually. \\
Thermocouples are another temperature monitoring option, which is widely used if sensing is required due to their simplicity and reliability \cite{Machin2018}. Thermocouples are relatively easy to fabricate and are self powered making them an ideal candidate for low-cost thermometry, since their signal stems from intrinsic material properties, they tend to have only minimal variations in sensitivity. As the name suggests, a thermocouple, in the classical sense, is typically a combination of two materials (often metals) with different Seebeck coefficients $S = -\Delta V / \Delta T$, which are joined at the sensing end \cite{Kinzie1973}. Then a temperature difference between the sensing end at $T_\mathrm{sense}$ and the measuring end at $T_\mathrm{meas}$ leads to the buildup of a thermovoltage via the Seebeck effect:
\begin{equation}
	V_\mathrm{th} = -[(S_{1}-S_{2})( T_\mathrm{sense}- T_\mathrm{meas})]~, 
	\label{EQ1:ThermocoupleDef}
\end{equation} 
where $V_\mathrm{th}$ refers to the thermovoltage drop across the two leads at the measuring end and $S_{1}$ and $S_{2}$ are the Seebeck coefficients of the two materials used. Depending on the desired working temperature range and the required sensitivity different material combinations are used. A multitude of different types of thermocouples has been developed, covering a wide range of temperatures and work environments \cite{Kazemi2017,Asamoto1967,Tougas2013,Pollock1991}. Typically, in order to achieve on-chip thermometry with conventional thermocouples, two separate fabrication runs are required. \\ 
It has previously been shown that it is possible to produce a single-material thermocouple by varying the width of thin gold stripes \cite{Sun2011,Liu2012} and other metals \cite{Szakmany2014}. The proposed mechanism is that reducing the width of the gold stripes changes the Seebeck coefficient due to increased scattering at the grain boundaries and structural defects. However, the sensitivity of all-metal thermocouples is only on the order of $1 \mathrm{\mu}$V/K and they tend to have a large footprint of tens of $\mathrm{\mu}$m in width and hundreds of $\mathrm{\mu}$m in length, which is too big for nanoelectronic applications. 
Here, we report the fabrication of two-dimensional thermocouples made out of single layer graphene. To this end, we make use of our recent discovery that similar to metals, the Seebeck coefficient in graphene can be influenced by geometrical constraint, which changes the mean free path locally \cite{Harzheim2018}. The advantage of using graphene compared to previous approaches is its high bulk Seebeck coefficient \cite{Zuev2009} and its long electron mean free path even at room temperature \cite{Banszerus2016}. This allows for the fabrication of highly sensitive thermocouples with the possibility of wafer-scale integration, as current research efforts are directed at the use of graphene in 2D van der Waals structures using wafer-scale fabrication methods \cite{Zhang2014,Avsar2011}. \\
\\
The graphene thermocouples are fabricated by patterning a U-shape into CVD-grown graphene consisting of a wide and a narrow leg.  To test the functionality of the thermocouples, an on-chip microheater is used to increase the temperature of the sensing end of the thermocouple (see Fig. \ref{Fig:1:Schematics}a). We fabricate both long ($L_1 = 4.3 \, \mu$m) and short ($L_2 = 2.5 \, \mu$m) thermocouples and vary the width of the narrow leg $w$ from $1 \, \mu$m to $0.2 \, \mu$m while keeping the width of the wide leg constant at $w_0 = 1.5 \, \mu$m. The sensing end of the thermocouple where the wide and narrow leg meet is located 700 nm from the heater.
\\
We use a Scanning Thermal Microscope (SThM) to calibrate the heater and quantify the temperature distribution along the device. To this end we performed an SThM measurement of the device structure at different heater currents and obtain a temperature profile map of the device (see Fig. \ref{Fig:1:Schematics}b). We used a method that eliminates the influence of fluctuations in the tip sample thermal resistance \cite{Menges2016,Konemann2019}. This enabled us to develop a model for the temperature gradient along the substrate and subsequently calculate the temperature difference $\Delta T = T_\mathrm{sense}-T_\mathrm{meas}$ (see Fig. \ref{Fig:1:Schematics}c and SI).

\begin{figure}
	\includegraphics[width=\linewidth]{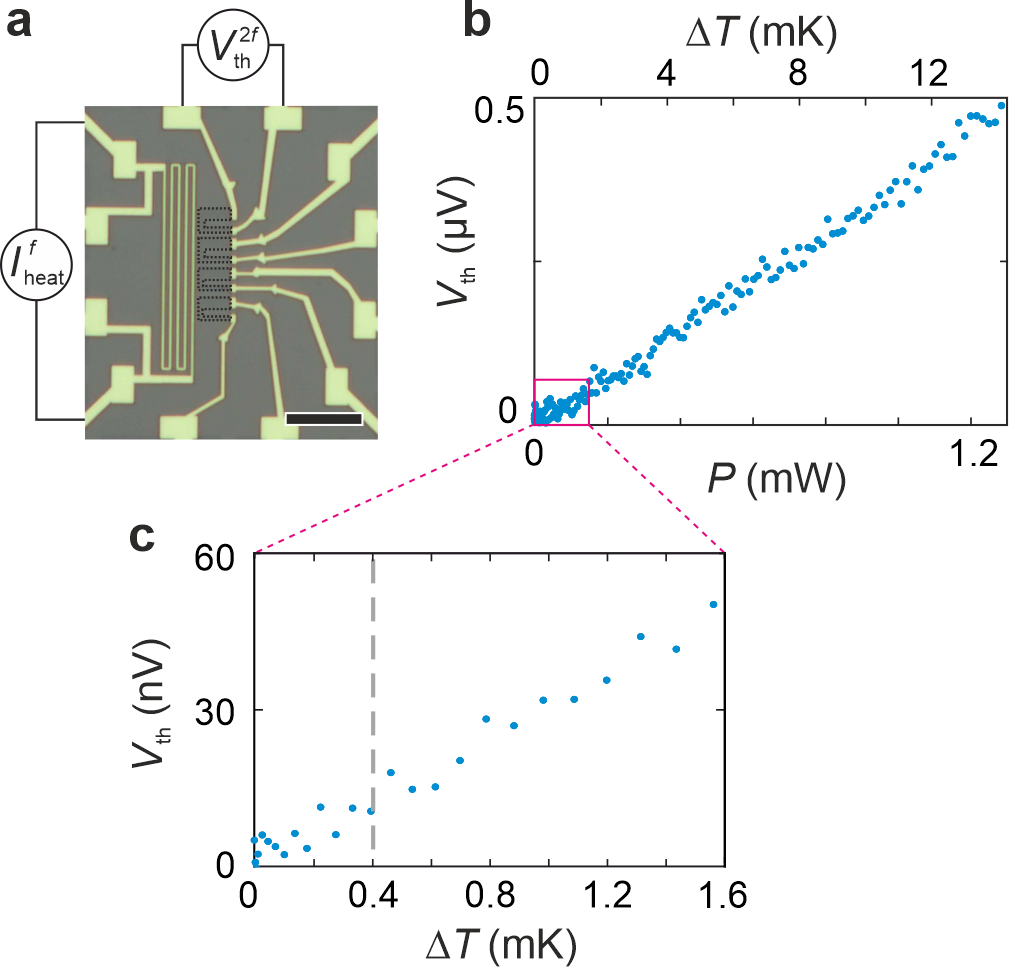}
	\caption{\label{Fig:2:Example Device}a) Optical microscope image of a typical device showing the measurement configuration. The graphene thermocouples next to the heater are highlighted by the black dotted lines and the scale bar denotes 10 $\mu$m. b) Thermovoltage response of a short device with $w = 0.2 \, \mu$m to heater power, the blow-up below shows the same measurement with higher integration time where the grey dashed line indicates the minimum temperature sensitivity.}	
\end{figure}

In order to quantify the sensitivity of the thermocouple, the heater is excited with a sinusoidal current at a frequency $f$ and the thermovoltage response of the thermocouple is recorded at a frequency $2f$ between the wide and narrow leg (see Fig. \ref{Fig:2:Example Device}a). For the purpose of this paper, all signals reported are the phase independent root-mean-square (RMS) values though the thermovoltage signal is mainly present at a phase of 90$^{\circ}$ with respect to the excitation \cite{Gehring2019}. Since we aim to determine the peak value of the thermovoltage response, the measured RMS signal is multiplied by a factor of $\sqrt{2}$. The resistance of the devices is extracted from a DC \textit{IV} trace, with all measured devices exhibiting a linear behaviour (see SI). Measurements are performed at room temperature and in a vacuum environment to prevent parasitic heat transfer due to conduction and convection.\\

\begin{figure}
	\includegraphics[width=\linewidth]{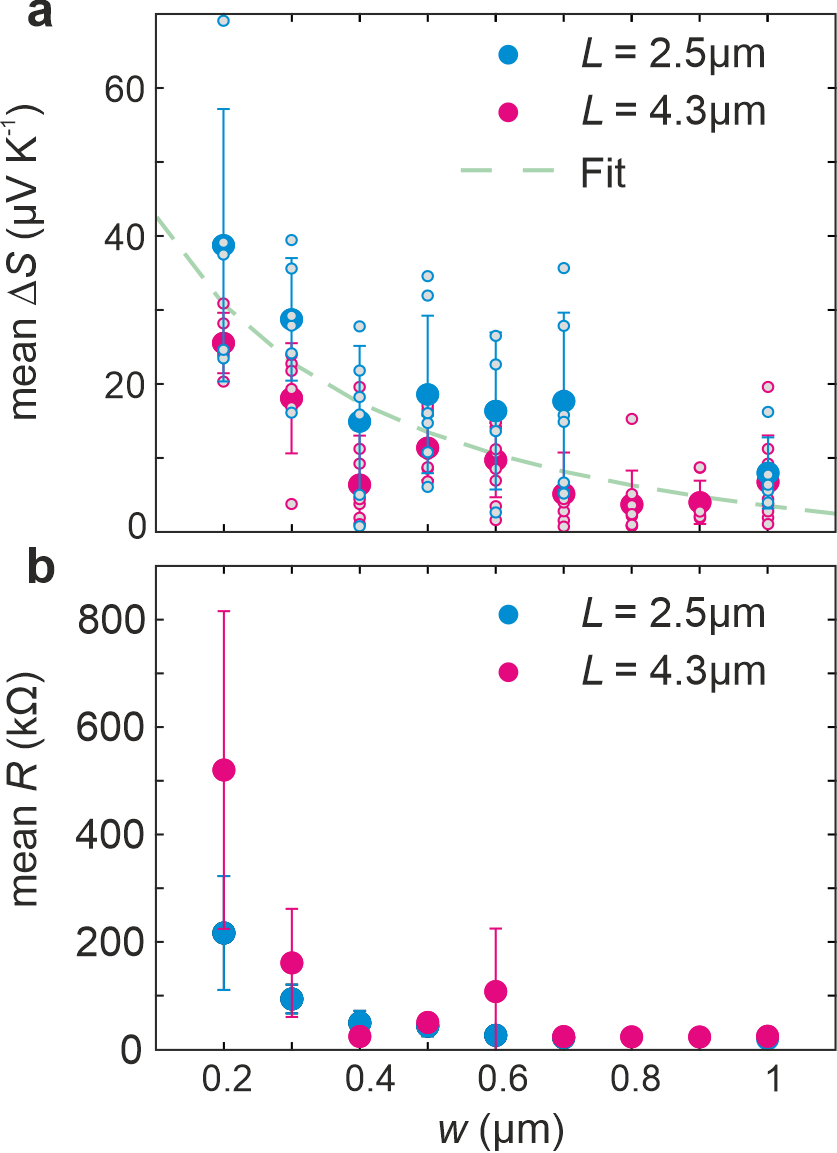}
	\caption{\label{Fig:3:Statistics}a) Mean $\Delta S$ signal as a function of the narrow leg width $w$ for $L = 2.5 \, \mu$m (blue dots) and $L = 4.3 \, \mu$m long (red dots) devices respectively. The grey dots indicate single device measurements for the long (red outline) and short devices (blue outline) respectively and the errorbars indicate the standard deviation with respect to these single measurements. The green dashed line is the fit using Equation \protect\ref{EQ4:ThermovoltageFit}. For the short length devices, the widths $w = 0.9 \, \mathrm{\mathrm{\mu}}$m and $w = $0.8 $\mathrm{\mathrm{\mu}}$m could not be measured due to a fabrication error. b) Average device resistance for the short (blue dots) and long (red dots) devices as a function of the narrow leg width $w$, where the errorbars indicate the standard deviation.}	
\end{figure}

As is shown in Fig. 2b, for the best performing device geometry, which is a short junction with a narrow leg width of $w = 0.2\, \mu$m (see analysis below), the junction responds linearly to the heater power $P$. Assuming that the temperature difference is proportional to the heater power, the observed behaviour is in agreement with the Joule-Lenz law which predicts a quadratic relation between the heating Power $P$ and the applied current $I$, $\Delta T = T_\mathrm{sense}- T_\mathrm{meas} \propto P = I^2R$. A higher heater power will induce a higher temperature difference $\Delta T$ which results in a higher thermovoltage signal, see equation (\ref{EQ1:ThermocoupleDef}). Using the SThM calibration allows us to convert the heater power $P$ into a corresponding temperature drop $\Delta T$ on the junction, which we show as a top x-axis in Fig. \ref{Fig:2:Example Device}b. \\
A clear increase in the thermovoltage signal is seen for relatively small temperature differences of a few mK. For a longer integration time (10 seconds compared to 1 second for Fig. \ref{Fig:2:Example Device}b) it is possible to push the response threshold to an even smaller temperature difference of $\Delta T \approx 400$ $\mu$K (see Fig. \ref{Fig:2:Example Device}c grey dashed line) highlighting the excellent sensing abilities of the graphene thermocouples: a maximum sensitivity of $\Delta S \approx$39 $\mathrm{\mathrm{\mu}}$V/K is reached for a width of $w=0.2 \, \mu$m and a length of $L=2.5 \, \mu$m. \\
The origin of the signal can be explained by a change in the Seebeck coefficient ($S_\mathrm{w_0}$ to $S_w$) when varying the leg width from $w_0 = 1.5 \, \mu$m to a narrower width $w$ \cite{Harzheim2018}. This change in the Seebeck coefficient is due to the increased influence of scattering from the edges on the mean free path as the width of the channel is reduced. As scattering is more prominent in the narrow channel the mean free path decreases. Here, the defect potentials responsible for the scattering stems from irregularities at the edges in the graphene devices, introduced by the oxygen plasma etching necessary to pattern the devices as well as atmospheric contamination.\\
Following a previously developed theory for CVD graphene in the diffusive transport regime \cite{Harzheim2018} it is then possible to arrive at a width dependent expression for the mean free path 
\begin{equation}
	l(w) = l_0 \bigg[1+c_n\bigg(\frac{l_0}{w}\bigg)^n\bigg]^{-1}~,
	\label{EQ2:Meanfreepath}
\end{equation}
where $l_0$ is the bulk mean free path and $c_n$ and $n$ are numerical coefficients specifying the transport mode and the influence of scattering on the mean free path. We can then combine equation (\ref{EQ2:Meanfreepath}) with the Mott formula for the Seebeck coefficient given by $S= \frac{\pi^2k_B^2T}{3 e}\frac{1}{R(\epsilon)}\frac{dR(\epsilon)}{d \epsilon}|_{\epsilon = \epsilon_F}$, where $R(\epsilon)$ is the energy dependent resistance in graphene. While the Seebeck coefficient of graphene can deviate from the Mott formula at high temperatures, the Mott formula gives a good indication of the signal and has been shown to reproduce the trend in experimental data correctly \cite{Zuev2009}. Using the Mott formula results in an expression for the width dependent Seebeck coefficient in graphene:
\begin{equation}
	S = -\frac{\pi^2k_B^2T}{3\epsilon_F e}\bigg[1+nU\frac{l(w)}{l_0}-(n-1)U \bigg]~.
	\label{EQ3:SeebeckwDependent}
\end{equation}
Here, $U = \frac{d \, ln (l_0)}{d \, ln(\epsilon)}\bigg|_{(\epsilon=\epsilon_F)}$ is the exponent of the power law dependence of the electron mean free path on energy and $\epsilon_F = \hbar v_F\sqrt{4 \pi n/g_sg_v}$ is the Fermi energy with the Fermi velocity $v_F = 10^6$ m/s, the spin and valley degeneracy $g_s = g_v = 2$ and $n = 10^{16}$$m^{-2}$ being the carrier density \cite{DasSarma2011}. Equation (\ref{EQ3:SeebeckwDependent}) predicts a decrease in thermopower as the width of the graphene channel is reduced. This means that the difference $\Delta S$ between $S_{w_0}$ and $S_{w}$ (see Figure \ref{Fig:1:Schematics}a) is expected to be largest for narrow $w$. The Seebeck coefficient difference $\Delta S$ can thus be expressed as
\begin{equation}
	\Delta S = (S_{w_0}-S_w) = - \frac{\pi^2k_B^2 T nU}{3 \epsilon_F e} \bigg[\frac{l(w_0)-l(w)}{l_0}\bigg] ~.
	\label{EQ4:ThermovoltageFit}
\end{equation}
Usually the Seebeck coefficient $S$ is temperature dependent, however for the small temperature gradients used in our study we can assume that $S_w(T_\mathrm{sense}) \approx S_w(T_\mathrm{meas})$ and similarly for $S_{w_0}$.\\ 
The width dependence of both the $\Delta S$ signal and the resistance for over 110 devices with differing widths $w$ from $1 \, \mathrm{\mathrm{\mu}}$m to $0.2 \, \mathrm{\mathrm{\mu}}$m is shown in Fig. \ref{Fig:3:Statistics} for the two different device lengths $L_1$ and $L_2$. The thermovoltage was recorded at a heater power of $P \approx 1.3$ mW for all devices. This corresponds to a temperature difference $\Delta T$ between the sensing end and the measuring end of $\Delta T_\mathrm{short} = 14.3 \pm 2.17$ mK for the short devices and $\Delta T_\mathrm{long} = 22.35 \pm 2.97$ mK for the long devices. As the narrow leg width decreases, the mean $\Delta S$ increases, in accordance with equation (\ref{EQ4:ThermovoltageFit}), which predicts that a lower mean free path in the narrow leg will lead to a higher Seebeck difference $\Delta S$ (Fig. \ref{Fig:3:Statistics}a). Similarly, the resistance of both the long and short devices increases with decreasing width $w$ (Fig. \ref{Fig:3:Statistics}b).\\

The increase in resistance (of Fig. \ref{Fig:3:Statistics}b) can be explained by the same formalism used to describe the thermoelectric properties of the system, where the CVD graphene on SiO$_2$ is treated as a diffusive conductor \cite{Adam2007}.
In the diffusive transport regime the conductance is proportional to the electron mean free path, $\sigma \propto l_\mathrm{e}$. Since $\sigma \propto 1/R$ we then expect $l_\mathrm{e} \propto 1/R$, meaning an increase in resistance for a smaller electron mean free path. This can be seen in Fig. \ref{Fig:3:Statistics}b where a narrowing leg width $w$ reduces the mean free path due to scattering along the rough edges resulting in a higher resistance $R$. \\
As the mean free path is the crucial factor in determining the resistance and equally the determining factor for $\Delta S$ (see equation \ref{EQ4:ThermovoltageFit}), a decrease in the resistance should simultaneously show up as a decrease in the $\Delta S$ signal and vice versa. This is evident in Fig. \ref{Fig:3:Statistics}a and b for the long devices, where small variations in the device resistance when changing $w$ from 0.5 to 0.4 $\mu$m are directly reflected in corresponding variations of the $\Delta S$ signal.\\

Both the short, $L=2.5 \, \mu$m, and the long, $L = 4.3 \, \mu$m devices show a similar trend of an increasing $\Delta S$ signal with a decreasing channel width (Fig. \ref{Fig:3:Statistics}a). The $\Delta S$ signal for both device lengths can be fitted using equation (\ref{EQ4:ThermovoltageFit}), accounting for the different $\Delta T$ and using the same fitting parameters for both lengths (see Fig. \ref{Fig:3:Statistics}a black dashed line). We find $c_n = 0.96$, $l_0 = 248$ nm, $n = 1.08$ and $U = 0.99$, which is similar to previously found values from measurements of the thermopower width dependence in CVD grown graphene \cite{Harzheim2018}. The $U$ value of approximately 1 points towards long-range Coulomb interaction being the determining factor in the mean free path, consistent with scattering centres along the graphene edges \cite{Castro2009,Li2013}. Differences in signal strength between the long and short devices, as well as for individual devices (see Fig. \ref{Fig:3:Statistics}a), can be attributed to the unpredictable nature of the defects in the narrow graphene legs, ultimately determining the size of $S_w$. This leads to a relatively high standard deviation for the $\Delta S$ signals in the graphene thermocouples, however we note that the Seebeck response of the short and long devices are within one standard deviation of each other.  \\

It should be noted that the critical dimensions of the graphene thermocouples are $w_0$ and $w$. Ideally $w_0$ should be larger than the mean-free path in the graphene to avoid influence from edge scattering, while $w$ should be as small as possible to maximize edge scattering. The minimum size of $w = 0.2 \, \mu$m in this study is due to limitations in our fabrication process.
Furthermore, an avenue to achieve more sensitive thermocouples is to increase the 'bulk' value $S$ of graphene, which can be achieved by oxygen plasma treatment of graphene \cite{Xiao2011}, operating in the hydrodynamic regime \cite{Ghahari2016}, encapsulating graphene in hBN \cite{Duan2016,Banszerus2016} as well as changing the carrier concentration and band structure through gating \cite{Zuev2009,Wang2011}. Another possible path is to use exfoliated graphene which has a higher electron mean free path and therefore a higher Seebeck coefficient than CVD grown graphene \cite{Novoselov2012} and in addition exhibits less defects than CVD graphene, making it easier controllable. However, similar to most of the other enhancement approaches mentioned above, the drawbacks are limited scalability of the graphene thermocouples due to a progressively more complex fabrication process. Nonetheless, an improved control over the edge configuration in graphene and therefore edge scattering and the mean free path is needed and further advances as well as more sophisticated fabrication methods should enable the creation of reproducible and well-defined thermocouples \cite{He2014,Zhang2013}.

In summary, we developed a single material thermocouple consisting of a U-shaped graphene structure with a narrow and a wide leg joined at the temperature sensing end. The behaviour of the thermocouple is well modelled by a previously developed theory on the dependence of the thermoelectric properties of graphene on its geometry. Furthermore, we demonstrated a higher sensitivity than in previously reported single material thermocouples by well over an order of magnitude at a footprint of only a few $\mathrm{\mathrm{\mu}}$m in width and length. The devices presented in this study thus could be used to facilitate cheap and easy to fabricate on-chip thermometry while being compatible with van der Waals heterostructures, current MOS-FETs and future graphene circuits. In addition, due to the bio-inert nature of graphene \cite{Shao2010}, as well as the small footprint and sub-millikelvin sensitivity of the graphene thermocouples they lend themselves to the increasingly relevant task of temperature probing of cells and other living systems \cite{Kucsko2013,Bai2016}.

\begin{acknowledgments}
This work was supported by the EC H2020 FET Open project 767187 QuIET and the EU Horizon 2020 research and innovation programme under grant agreement No. 785219 Graphene Flagship. P.G. acknowledges a Marie Skodowska-Curie Individual Fellowships (Grant No. TherSpinMol-748642) from the European Unions Horizon 2020 research and innovation programme.
\end{acknowledgments}
\vspace{5mm}
\textbf{References}
%

\clearpage

\onecolumngrid

\setcounter{equation}{0}
\setcounter{figure}{0}
\renewcommand{\theequation}{S\arabic{equation}}
\renewcommand{\thefigure}{S\arabic{figure}}

\large{\textbf{SI: All-graphene thermocouples}}
\section{Fabrication}
The devices were fabricated by transferring CVD grown graphene on top of a Si chip with a 300 nm SiO$_2$ layer. Subsequently, the contacts to the graphene and the heater are patterned via electron beam lithography and the Ti/Au contact and heater structure is evaporated on top of the graphene. In a last step, the device structure is defined and the graphene is etched away via oxygen plasma etching, leaving us with a U-shaped graphene thermocouple. The thermocouples are either 2.5 or 4.3 $\mu$m long and the sensing end, i.e. the point where the wide leg joins the narrow leg, is located 700 nm from the heater (see Fig. \ref{Fig:1:Schematics}).
\section{Resistance measurement}
\begin{figure} [h]
	\includegraphics[width=0.5\linewidth]{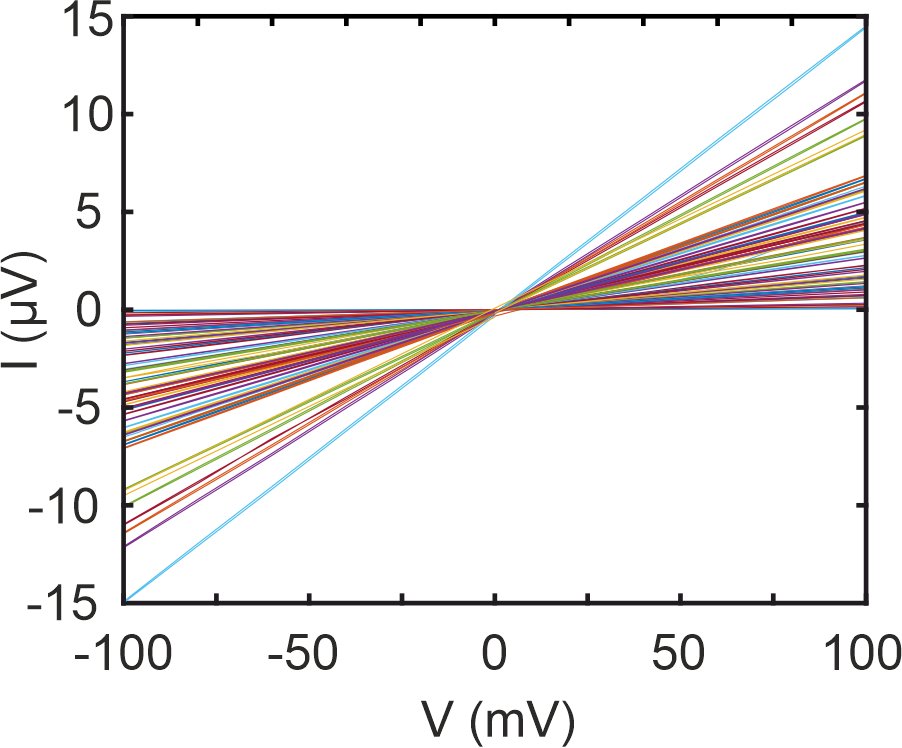} 
	\caption{\label{Fig:SI:R} IV traces for a variation of graphene thermocouples displaying the linear current dependencies of the devices.}	
\end{figure}

For all devices, an IV trace was recorded prior to the thermovoltage measurement, confirming that the devices were alive. All live devices showed a linear current dependency and were subsequently fitted to extract the resistance $R$ (see Fig. \ref{Fig:SI:R}). In addition, to ensure the device is not connected to the heater or the devices to each other, the resistance between neighbouring devices and between devices and the heater meander structure was tested.

\section{SThM measurement}
The temperature calibration is realized using an SThM based system utilising a micro electro mechanical system (MEMS) cantilever which consists of an integrated resistive sensor coupled to a Si tip $^1$. For the calibration the sensor is used in active mode, that means a voltage is applied to the cantilever, resulting in a significant temperature rise for an out of contact tip $^2$.\\
The calibration measurement is performed in a low-noise environment and a high vacuum at room temperature. The tip is brought into contact with the sample and scanned across the respective device. At the same time, the microheater next to the graphene thermocouples is excited with an AC current, modulating the heater temperature. The resulting temperature increase of the device and the substrate can then be calculated from a simultaneous measurement of the time averaged sensor signal $\Delta V_\mathrm{DC}$ and the demodulated sensor signal amplitude at the second harmonic $\Delta V_\mathrm{AC}$ 

\begin{equation*}
\Delta T = \Delta \phi \frac{\Delta V_{AC}}{\Delta V_{AC}-\Delta V_{DC}}~.
\end{equation*}
Here $\Delta \phi$ is the temperature of the sensor in the cantilever as a result of the applied bias. 
In order to ensure a constant contact force, the cantilever is monitored and controlled via a laser deflection system. The main advantage of this method is that fluctuations in the tip-sample thermal contact resistance, that can lead to substantial artifacts, are excluded in deriving the equation above as reported previously $^2$.

\section{Temperature calibration}
The temperature gradient along the device is determined as follows. For multiple different heater powers, an SThM map of a device is recorded (see Fig. \ref{Fig:1:Schematics}). Subsequently, a line cut perpendicular to the heater is extracted (black dashed line in Fig. \ref{Fig:1:Schematics}b and red line in Fig. \ref{Fig:1:Schematics}c) which shows the temperature distribution along the substrate. \\
The temperature drop over the substrate induced by the heater can be fitted with an exponential model, $dT = aI^2e^{b*x}$ where $a$ and $b$ are fitting parameters, $I$ is the heater current and $x$ is the distance along the line. We fit only the part of the substrate at least 500 nm away from the heater as can be seen in Fig. \ref{Fig:1:Schematics}c to only take into account the substrate effects. From the model and the  position of the device (starting 700 nm away from the heater and ending 4.3 $\mu$m or 5 $\mu$m from it for the short or long devices respectively) which is inferred from the known device dimensions and the topography map we can then determine the temperature at the respective sensing points. Since we also know the power dissipated in the heater we can calculate the power and distance dependent temperature along the substrate in the units $K/m/W$. Plugging in the respective device lengths and the power through the heater during measurements then gives $\Delta T_\mathrm{long} = 22.35 \pm 2.97$ mK and $\Delta T_\mathrm{short} = 14.3 \pm 2.17$ mK for the long and short devices respectively. These values were calculated for multiple calibrated devices and are all within one standard deviation of each other. Since the error introduced by the device variability is much bigger than the standard deviation in the temperature, the error in $\Delta T$ is neglected for the analysis of the results. 

\vspace{5mm}
\textbf{References}\\

\noindent \small $^1$F. Könemann, M. Vollmann, T. Wagner, N. Mohd Ghazali, T. Yamaguchi,
A. Stemmer, K. Ishibashi, and B. Gotsmann, “Thermal Conductivity of a
Supported Multiwalled Carbon Nanotube,” Journal of Physical Chemistry
C \textbf{123}, 12460–12465 (2019).\\
\small $^2$Menges, P. Mensch, H. Schmid, H. Riel, A. Stemmer, and B. Gotsmann,
“Temperature mapping of operating nanoscale devices by scanning probe
thermometry,” Nature Communications \textbf{7}, 10874 (2016).
\end{document}